\documentclass[twocolumn,prb,showpacs,preprintnumbers,amsmath,amssymb,floatfix]{revtex4}
\usepackage{graphicx}
\usepackage{amsmath}
\usepackage{amssymb}
\usepackage{color}

\begin{document}
%

%title and authors
\title{Electron-phonon effects on spin-orbit split bands of two dimensional systems}

\author{E. Cappelluti$^{1,2}$}
\author{C. Grimaldi$^{3,4}$}
\author{F. Marsiglio$^{5}$}

\affiliation{$^1$SMC Research Center, INFM-CNR c/o Dept. Physics,
University ``La Sapienza'', P.le  A. Moro 2, 00185 Roma, Italy}

\affiliation{$^2$Istituto dei Sistemi Complessi (ISC), CNR, v. dei
Taurini 19, 00185 Roma, Italy}

\affiliation{$^3$ Max-Plank-Institut f\"ur Physik komplexer Systeme,
N\"othnitzer Srt.38, D-01187 Dresden Germany}

\affiliation{$^4$ LPM, Ecole Polytechnique F\'ed\'erale de
Lausanne, Station 17, CH-1015 Lausanne, Switzerland}

\affiliation{$^5$Department of Physics, University of Alberta,
Edmonton, Alberta, Canada, T6G 2J1}

%\date{15-09-04}

%\widetext

\begin{abstract}
The electronic self-energy is studied for a two dimensional electron
gas coupled to a spin-orbit Rashba field and interacting with
dispersionless phonons. For the case of a momentum independent
electron-phonon coupling (Holstein model) we solve numerically the
self-consistent non-crossing approximation for the self-energy and
calculate the electron mass enhancement $m^*/m$ and the spectral
properties. We find that, even for nominal weak electron-phonon
interaction, for strong spin-orbit couplings the electrons behave as
effectively strongly coupled to the phonons. We interpret this
result by a topological change of the Fermi surface occurring at
sufficiently strong spin-orbit coupling, which induces a square-root
divergence in the electronic density of states at low energies. We
provide results for $m^*/m$ and for the density of states of the
interacting electrons for several values of the electron filling and
of the spin-orbit interaction.
\end{abstract}
\maketitle

\section{Introduction}
\label{intro}

Prompted by considerable technological interests, the physics of
itinerant electrons coupled to spin-orbit (SO) potentials has been
the subject of extensive investigations in recent
years.\cite{prinz,fabian} In materials of interest, the main sources
of  SO coupling are the Rashba interaction arising from structural
inversion asymmetries of low-dimensional structures,\cite{rashba}
and the Dresselhaus interaction present in bulk crystals lacking
inversion symmetry.\cite{dressel} Depending on the material
characteristics, one of the above interactions, or even both, may be
present, lifting the spin degeneracy of the electron dispersion.
When measured at the Fermi level, the resulting energy splitting,
$\Delta_{\rm so}$, is commonly used to estimate the strength of the SO
interaction.

In narrow-gap III-V semiconductor-based heterostructures, such as
GaAs and InAs quantum wells, $\Delta_{\rm so}$ is a few meV, while
in II-VI quantum wells $\Delta_{\rm so}$ is greatly enhanced. For
example the heavy-hole conduction band of HgTe displays SO splitting
values ranging between $10-17$ meV and $30$ meV.\cite{zhang,gui}
Much stronger SO splittings have been observed in the surface states
of metals\cite{lashell} and semimetals\cite{koroteev,sugawara}, and
the corresponding $\Delta_{\rm so}$ may be so large, {\it e.g.}
$\Delta_{\rm so}\simeq 110$ meV in Au(111),\cite{lashell} that the
possibility of detecting SO split image states has been recently
put forward.\cite{mclaughlan} Other systems displaying giant SO
splittings are surface alloys as, for example,
Li/W(110),\cite{rotenberg} Pb/Ag(111),\cite{pacile,ast1} and
Bi/Ag(111),\cite{astprl} or even one-dimensional structures such as Au
chains in vicinal Si(111) surfaces.\cite{barke} For such
low-dimensional or structured materials, the SO interaction is of
Rashba type, but large SO splittings have been found (or predicted)
also in bulk crystals, where the Dresselhaus interaction leads to
$\Delta_{\rm so}$ as large as $200$ meV in non-centrosymmetric
superconductors CePt$_3$Si,\cite{bauer,samokhin} Li$_2$Pd$_3$B, and
Li$_2$P7$_3$B.\cite{togano,yuan}

Such strong SO couplings may possibly have interesting applications
in spintronic devices, but represent also a compelling and challenging
problem from the theoretical standpoint, in particular when
$\Delta_{\rm so}$ is no longer the smallest energy scale in the system,
as in III-V semiconductor heterostuctures where
$\Delta_{\rm so}\approx 1-5$ meV, but competes in magnitude with other
characteristic energy scales such as the phonon frequency or the Fermi energy.
From this perspective, systems like the Bi/Ag(111) surface alloy, which
shows bands split by about $200$ meV,\cite{ast1,astprl} are particularly promising,
given also the alleged possibility of tuning, by Pb doping, the Fermi energy
$E_F$ to values lower than the SO energy splitting.\cite{astold}

A few novel and interesting features arising from strong SO
splittings have already been investigated theoretically in the
literature. For example, in Ref.[\onlinecite{chaplik}] it has been
demonstrated that the Rashba SO coupling induces an infinite number
of bound states in two dimensions, even for short ranged impurity
potentials, while in a recent work we have shown that the
superconducting critical temperature of a low-density
two-dimensional (2D) electron gas can be significantly enhanced by
the Rashba interaction.\cite{cgm2007} Both phenomena discussed in
Refs.[\onlinecite{chaplik,cgm2007}] can be understood in terms of a
SO induced topological change of the Fermi surface, which gives rise
to an effective reduction of dimensionality of the electronic
density of states for $E_F$ sufficiently smaller than the SO
characteristic energy.

In this paper we analyze the effects of such topological change of
the Fermi surface on the electron-phonon (el-ph) problem of 2D
systems. In particular, we study one-particle spectral properties and
extract the combined el-ph and SO effects on the electronic
effective mass $m^*$ and on the interacting density of states (DOS).
We show that, even for weak or moderate couplings to phonons, the
effective reduction of the bare DOS induced by the Rashba
interaction leads to a strong increase of $m^*$, and to phonon
satellite peaks in the interacting DOS, which are typical signatures
of an effectively strong el-ph coupling. Due to the
two-dimensionality of our model, and to the Rashba type of SO
coupling, our results could be relevant for both metal and semimetal
surface states, for which the el-ph interaction has been shown to be
relevant,\cite{sugawara,gayone,hofmann,lashell2,kroger} and for
surface superconductors,\cite{gorkov} with the hypothesis that
pairing is provided by the coupling to phonons.

\section{Rashba-Holstein model}
\label{model}

\begin{figure}
\protect
\includegraphics[scale=0.45,clip=true]{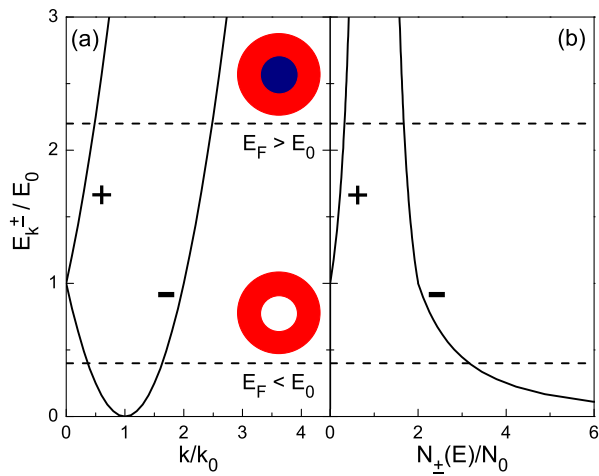}
\caption{(color online). (a): electron dispersion
$E_k^{\pm}=\frac{\hbar^2}{2m}(k\pm k_0)^2$ for a spin-orbit split
electron gas. The energy $E_0=\frac{m}{2\hbar^2}\gamma^2$ is a
measure of the spin-orbit interaction and is equivalent to the
minimum interband excitation energy for an electron sitting at
the bottom of the lower band. The upper and lower horizontal dashed lines
indicate the position of the Fermi level for $E_F > E_0$
and $E_F < E_0$, respectively. Also shown are the corresponding
Fermi circles with occupied states drawn by gray (colored) regions.
(b): density of states plotted from
Eqs.(\ref{DOS1},\ref{DOS2}). } \label{fig1}
\end{figure}

Two-dimensional quantum wells, with strong and asymmetric confining
potentials, and surface states with weak or negligible coupling to
the bulk can be satisfactorily represented by the following 2D
electron hamiltonian with SO interaction
\begin{equation}
\label{ham}
H_0=\sum_{\mathbf{k}\alpha}\epsilon_{\mathbf{k}}c^\dagger_{\mathbf{k}\alpha}
c_{\mathbf{k}\alpha}+\sum_{\mathbf{k}\alpha\beta}\mathbf{\Omega}_{\mathbf{k}}
\cdot\mbox{\boldmath$\sigma$}_{\alpha\beta}c^\dagger_{\mathbf{k}\alpha}
c_{\mathbf{k}\beta}
\end{equation}
where $c^\dagger_{{\bf k}s}$ ($c_{{\bf k}s}$) is the creation (annihilation)
operator for an electron with momentum ${\bf k}=(k_x,k_y)$ and spin index
$\alpha=\uparrow,\downarrow$. In the above expression,
$\epsilon_{{\bf k}}$ is the electron dispersion in the absence of SO
coupling, $\mbox{\boldmath$\sigma$}$ is the spin-vector operator with components
($\sigma^x,\sigma^y,\sigma^z$) given by the Pauli matrices, and
$\mathbf{\Omega}_{\bf k}$ is a ${\bf k}$ dependent SO
pseudopotential arising from the asymmetry in the $z$-direction of the
confining potential.  Here we consider a linear Rashba model for the SO
interaction
\begin{equation}
\label{rashba1} \mathbf{\Omega}_{{\bf
k}}\cdot\mbox{\boldmath$\sigma$}=\gamma(k_x\sigma^y-k_y\sigma^x),
\end{equation}
where $\gamma$ is the SO coupling constant. Furthermore, we assume
that the unperturbed electron band is parabolic, $\epsilon_{{\bf
k}}=\hbar^2k^2/2m$, where $m$ is the band mass of the electron.
Apart from a constant shift $E_0$ (defined below) which can be
absorbed in the chemical potential, the eigenvalues of
Eqs.(\ref{ham}) and (\ref{rashba1}) are:
\begin{equation}
\label{energy2}
E_k^s=\frac{\hbar^2}{2m}(k +s k_0)^2,
\end{equation}
where $k=\vert{\bf k}\vert$, $s=\pm$ is the chirality number,
and $k_0$ is the Rashba momentum
\begin{equation}
\label{k0}
k_0=\frac{m}{\hbar^2}\gamma .
\end{equation}
The two electron branches $E_k^{\pm}$ are plotted in Fig. \ref{fig1}(a)
in units of the Rashba energy
\begin{equation}
\label{E0}
E_0=\frac{\hbar^2 k_0^2}{2m}=\frac{m}{2\hbar^2}\gamma^2
\end{equation}
which corresponds to the energy difference between the degeneracy
point at $k=0$ and the bottom of the lower band at $k=k_0$. In Fig.
\ref{fig1}(a) we indicate also the Fermi levels for the $E_F>E_0$
and $E_F<E_0$ cases (horizontal dashed lines) which represent two
qualitatively different situations. For $E_F>E_0$, the Fermi level
crosses bands of different chirality and the corresponding Fermi sea
is given by the area of two concentric Fermi circles, as sketched
in Fig. \ref{fig1}(a). In this case,
the corresponding DOS for each sub-band is
\begin{equation}
\label{DOS1}
N_\pm(E_F)=N_0\left(1\mp\sqrt{\frac{E_0}{E_F}}\right)\,\,\,\,\,\,{\rm for}\,\, E_F\ge E_0
\end{equation}
where $N_0=m/2\pi\hbar^2$ is the DOS per spin direction with zero SO
coupling. The sum over the two chiral states,
$N(E_F)=N_+(E_F)+N_-(E_F)$, is therefore identical to the total DOS,
$2N_0$, of a 2D electron gas without SO interaction [Fig.
\ref{fig1}(b)]. Furthermore, in the $E_F\gg E_0$ regime, one has
$N_\pm(E_F)\simeq N_0$, and the dispersions of the low excitations
in the vicinity of $E_F$ can be approximated by $v_F(k-k_F)\pm
\Delta_{\rm so}/2$, where $v_F$ and $k_F$ are respectively the Fermi
velocity and momentum in the absence of SO interaction and
$\Delta_{\rm so}=2\gamma k_F$ is the SO energy splitting. This is
the quantity which is usually used to quantify the SO strength in
semiconductors such as GaAs and InAs.

For $E_F < E_0$ the situation is drastically different. In this case
in fact, as shown in Fig. \ref{fig1}(a), the Fermi level crosses
only the $s=-1$ band but, since $E_k^-$ has a minimum at $k=k_0\neq
0$, the Fermi surface is still constituted by two concentric
circles. The resulting Fermi sea is therefore given by the area of
the annulus comprised by the two circles and in the limit of
$E_F\rightarrow 0$, with $E_0\neq 0$, the Fermi surface $S_F$
coalesces into a circle of radius $k_0$, $S_F=2\pi k_0$, while the
Fermi velocity $v_F$ vanishes as $\sqrt{E_F}$. Since $N(E_F)\propto
S_F/v_F$, the resulting DOS is therefore:\cite{cgm2007}
\begin{equation}
\label{DOS2}
N(E_F)=N_-(E_F)=2 N_0\sqrt{\frac{E_0}{E_F}}\,\,\,\,\,\,{\rm for}\,\, E_F < E_0.
\end{equation}
As we shall see in the following, the one-dimensional-like singularity
of Eq.(\ref{DOS2}) has important and peculiar effects on the low-energy
properties of the system, in contrast with the $E_F>E_0$ case, for which
the corresponding DOS is featureless.

Let us introduce now the coupling to the phononic degrees of
freedom. In the present paper, we consider the following
Holstein-type of interaction hamiltonian
\begin{equation}
\label{ham2}
H_{ph}= \sum_{\bf q}\omega_0a^\dagger_{\bf q}a_{\bf q}+
g\sum_{\mathbf{k}\mathbf{k}'\alpha}
c^\dagger_{\mathbf{k}\alpha}c_{\mathbf{k}'\alpha}
(a^\dagger_{\mathbf{k}-\mathbf{k}'}+a_{\mathbf{k}'-\mathbf{k}}),
\end{equation}
where $a^\dagger_{\bf q}$ ($a_{\bf q}$) is the creation
(annihilation) operator for a phonon with momentum ${\bf q}$,
$\omega_0$ is a dispersionless phonon frequency, and $g$ is the
momentum independent el-ph matrix element. As will become clear in
the following, the choice of the momentum independent quantities
$\omega_0$ and $g$ is convenient for the calculation of the
self-energy, and permits a more direct evaluation of the effects of
the SO interaction on the el-ph properties. The present analysis is
therefore a starting point for more general formulations of the
el-ph hamiltonian.

The thermal Green's function of the electrons subjected to the total
hamiltonian $H=H_0+H_{\rm ph}$ satisfies the following Dyson equation
\begin{equation}
\label{self1} \mathbf{G}({\bf
k},i\omega_n)=\left[\mathbf{G}_0^{-1}({\bf
k},i\omega_n)-\mathbf{\Sigma}({\bf k},i\omega_n)\right]^{-1},
\end{equation}
where $\omega_n=(2n+1)\pi T$ is a Fermionic Matsubara frequency and $T$ is
the temperature. $\mathbf{G}_0({\bf k},i\omega_n)$ is the non-interacting
electron propagator and $\mathbf{\Sigma}({\bf
k},i\omega_n)$ is the self-energy due to the coupling with phonons.
Due to the SO interaction appearing in $H_0$, these quantities are
$2\times 2$ matrices in the spin sub-space. From Eqs. (\ref{ham}) and
(\ref{rashba1}), the non-interacting propagator is
\begin{equation}
\label{green1} {\bf G}_0({\bf k},i\omega_n)=\frac{1}{2}\sum_{s=\pm
1}(1+s\hat{\mathbf{\Omega}}_{\bf
k}\cdot\mbox{\boldmath$\sigma$})G_0(E^s_k,i\omega_n),
\end{equation}
where $\hat{\mathbf{\Omega}}_{\bf
k}\cdot\mbox{\boldmath$\sigma$}=\hat{k}_x\sigma_y-\hat{k}_y\sigma_x$,
and $G_0(E^s_k,i\omega_n)=1/(i\omega_n-E^s_k+\mu),$ where $\mu$
is the chemical potential.

For the evaluation of the self-energy, we shall consider a self-consistent
Born approximation (non-crossing approximation) which neglects all
el-ph vertex corrections. Furthermore, we shall not consider
many-body corrections to the phonon propagator. These limitations will be
discussed in Sec.\ref{disc} and, for the moment, it suffices to
keep in mind that this approximation scheme should be not too poor as long as
the coupling to the phonons is sufficiently weak. Hence, given the
phonon propagator
\begin{equation}
\label{self2} D(i\omega_n-i\omega_m)=
\frac{\omega_0^2}{(i\omega_n-i\omega_m)^2-\omega_0^2},
\end{equation}
the resulting electron self-energy matrix in the non-crossing approximation
reduces to:
\begin{equation}
\label{self3} \mathbf{\Sigma}({\bf
k},i\omega_n)=
-\frac{\lambda}{N_0}T\sum_m\! \int\!\!\frac{d{\bf k}'}{(2\pi)^2}
D(i\omega_n-i\omega_m)\mathbf{G}({\bf k}',i\omega_m),
\end{equation}
where $\lambda=2g^2 N_0/\omega_0$ is the
el-ph coupling constant. From Eq.(\ref{self3}) it is clear that,
due to the momentum independence of the el-ph
interaction, the self-energy  (\ref{self3}) depends only upon the
frequency. Furthermore, by substituting ${\bf G}_0({\bf k}',i\omega_m)$
for ${\bf G}({\bf k}',i\omega_m)$ in Eq.(\ref{self3}), the resulting
second-order self-energy is diagonal in the spin space. This holds true
for all orders of iteration, so that $\mathbf{\Sigma}({\bf
k},i\omega_n)=\Sigma(i\omega_n)\mathbf{1}$, where $\mathbf{1}$ is the unit matrix.
The Green's function
(\ref{self1}) can therefore be rewritten as
\begin{equation}
\label{self4} {\bf G}({\bf k},i\omega_n)=\frac{1}{2}\sum_{s=\pm
1}(1+s\hat{\mathbf{\Omega}}_{\bf
k}\cdot\mbox{\boldmath$\sigma$})G(E^s_k,i\omega_n),
\end{equation}
where
\begin{equation}
\label{self5}
G(E^s_k,i\omega_n)=\frac{1}{i\omega_n-E^s_k+\mu-\Sigma(i\omega_n)},
\end{equation}
is the electron propagator in the chiral basis for the interacting
case while the self-energy is
\begin{equation}
\label{self6} \Sigma(i\omega_n)=-\lambda
T\!\sum_m\!D(i\omega_n-i\omega_m) g(i\omega_m),
\end{equation}
where
\begin{equation}
\label{self7} g(i\omega_m)=\frac{1}{2N_0}
\sum_s\!\int_0^{k_c}\!\frac{dk k}{2\pi}\, G(E^s_k,i\omega_m).
\end{equation}
In the above expression, we have introduced an upper momentum
cut-off $k_c$ which prevents the integral over $k$ from diverging.
Such divergence is an artifact due to the use of a momentum
independent el-ph matrix element $g$ in Eq.(\ref{ham2}) and of the
electron gas model of $H_0$. On physical grounds, the introduction
of $k_c$ is equivalent therefore to defining a finite Brillouin zone
of area $\pi k_c^2$ or, equivalently, a finite bandwidth
$E_c=\hbar^2k_c^2/2m$ when $E_0=0$. In the following, $E_c$ will be
chosen to be much larger than the other relevant energy scales of
the system ($E_c\gg\omega_0$, $E_0$, $E_F$). A finite $k_c$, or
$E_c$, permits also to define a finite electron density
$\rho_e=\sum_\sigma\int d{\bf k}/(2\pi)^2 \langle c^\dagger_{{\bf
k}\sigma}c_{{\bf k}\sigma}\rangle$ which, relative to the cut-off
$k_c$, becomes
\begin{eqnarray}
\label{enne1} \rho_e&=&\sum_s\int_0^{k_c}\!\frac{dk k}{2\pi}
T\sum_n G(E_k^s,i\omega_n)
e^{i\omega_n o^+} \nonumber \\
&=&\frac{k_c^2}{4\pi}+T\!\sum_n {\rm Re}\,g(i\omega_n),
\end{eqnarray}
where $o^+$ is an infinitesimal positive quantity and the second
equality has been obtained by using  $T\sum_n G(E_k^s,i\omega_n)
e^{i\omega_n o^+}=1/2+T\sum_n {\rm Re}
G(E_k^s,i\omega_n)$.\cite{abri}  In the following, we shall
present results in terms of the electron number density
\begin{equation}
\label{enne2} n_e=\frac{4\pi\rho_e}{k_c^2}
\end{equation}
which attains the limiting value $n_e=2$ ($n_e=0$) for
completely filled (empty) bands.

Before turning to the next sections, where we present our numerical
results, it is worthwhile showing how the SO effects on the DOS
enter the self-energy function. By transforming the integration over
$k$ in an integration over the energy, Eq.(\ref{self7}) can be
rewritten as follows:
\begin{equation}
\label{self8}
g(i\omega_m)=\int_0^{E_c}\!dE\,
\rho_0(E)G(E,i\omega_m),
\end{equation}
where, for simplicity, terms of order $\sqrt{E_0/E_c}$ have been omitted
in the upper limit of integration, and
\begin{equation}
\label{DOS} \rho_0(E)=\sum_s\frac{N_s(E)}{2N_0}=\left\{
\begin{array}{ccc}
1 & {\rm for} & E\ge E_0 \\
\sqrt{\frac{E_0}{E}} & {\rm for } & E< E_0
\end{array} \right. ,
\end{equation}
is the reduced non-interacting DOS obtained from Eqs.(\ref{DOS1}) and (\ref{DOS2}).
From the above expressions, it is therefore straightforward to realize
the importance on the el-ph properties of the square root
singularity of the DOS at low energies. As we shall see in the following,
the effective electron mass and the electron spectral properties in the
presence of SO interaction will differ qualitatively from the corresponding
results for $E_0=0$.

\section{effective mass}
\label{effmass}

\begin{figure}[t]
\protect
\includegraphics[scale=0.39,clip=true]{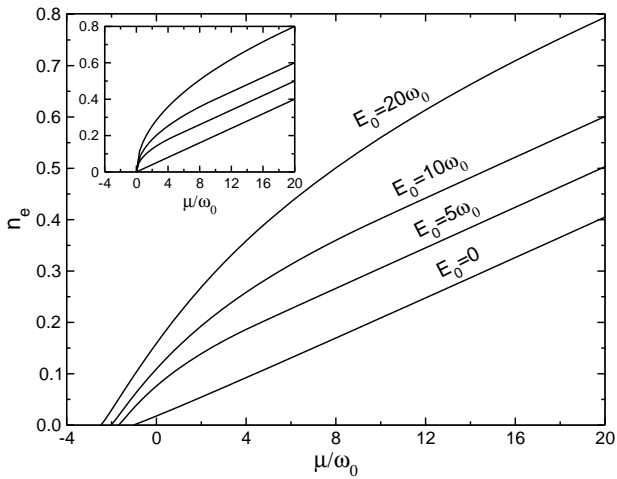}
\caption{Electron density number $n_e$ as a function of the
bare chemical potential $\mu$ for $\lambda=0.5$ and several values
of the SO interaction $E_0$. The temperature is $T=0.02\omega_0$ and
the energy cut-off is $E_c=100\omega_0$.
In the inset $n_e$ is plotted for $\lambda=0$ and for zero temperature.} \label{fig2}
\end{figure}

The integration over the momenta appearing in Eq.(\ref{self7}), or,
equivalently, the integration over the energy in Eq.(\ref{self8}),
can be carried out analytically, leaving only the summation over the
Matsubara frequency to be performed numerically. Hence, for fixed
values of $\lambda$, $\omega_0$ and $E_0$, the electron self-energy
$\Sigma(i\omega_n)$ is obtained by iteration of Eqs.(\ref{self5}),
(\ref{self6}), and (\ref{self7}), while Eq.(\ref{enne1}) is used to
extract the corresponding electron density for a given value of
$\mu$. For all cases we have set $E_c=100\omega_0$ and
$T=0.02\omega_0$, which is low enough to be representative of the
zero temperature case. In Fig. \ref{fig2} we report the calculated
values of $n_e$, Eq.(\ref{enne2}), for $\lambda=0.5$ and for
different values of the SO energy $E_0$. For comparison, we report in
the inset of Fig. \ref{fig2} the corresponding density values for $\lambda=0$
and at zero temperature. For $E_0=0$, $n_e$
decreases almost linearly as $\mu$ is reduced, as expected for a
constant DOS in 2D (see inset), but the zero density limit $n_e=0$
(extracted in the $T\rightarrow 0$ limit) is reached
only for $\mu=\mu_0\simeq -1.023 \omega_0$, which is lower than the
non-interacting zero-density value $\mu=0$. This energy decrease
represents the ground state energy of a single electron in
interaction with phonons and provides a measure of the strength of
the el-ph interaction. For non-zero SO coupling, $E_0>0$, two
features are apparent in Fig. \ref{fig2}. First, in the low density
limit, $n_e$ is no longer a linear function of $\mu$ and, second,
the ground state energy $\mu_0$ is even more lowered with respect to
the $E_0=0$ case. This latter feature indicates that, for fixed
$\lambda$, a single electron is more strongly coupled to phonons as
$E_0$ increases.

\begin{figure}[t]
\protect
\includegraphics[scale=0.39,clip=true]{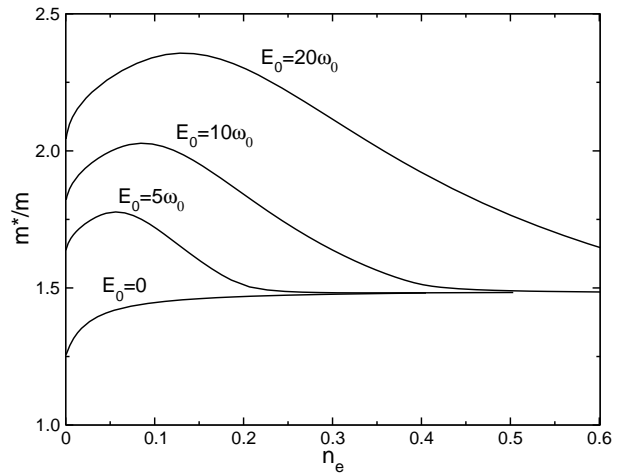}
\caption{Electron effective mass $m^*$ as a function of the electron density number $n_e$
for $\lambda=0.5$ and for several values of the SO interaction $E_0$.} \label{fig3}
\end{figure}

A more quantitative estimation of the role of SO coupling on the el-ph
properties is given by the electron effective mass enhancement $m^*/m$.
This quantity can be evaluated from
\begin{equation}
\label{mass}
\frac{m^*}{m}=1-\frac{{\rm Im}\Sigma(i\omega_n)}{\omega_n}\vert_{n=0}
\end{equation}
provided sufficiently low temperatures are considered. We have
checked that for $T=0.02\omega_0$ the effective mass ratio extracted
from Eq. (\ref{mass}) is in very good accord with the mass
enhancement obtained from the real frequency self-energy (see next
section). In Fig. \ref{fig3} we report our results for $m^*/m$ as a
function of the electron number density $n_e$ for the same parameter
values of Fig. \ref{fig2}. For $E_0=0$ we obtain the typical trend
for a 2D electron gas in the non-crossing approximation: $m^*/m$ is
almost a constant and approximately equal to the Migdal-Eliashberg
result $1+\lambda$ for relatively large densities while, for
$n_e\rightarrow 0$, $m^*/m$ decreases towards the one electron
result.\cite{ccgp} For $E_0=5\omega_0$, the mass enhancement follows
the $E_0=0$ case for densities larger than $n_e\simeq 0.2$,
corresponding to the range of densities for which $n_e$ is
proportional to $\mu$ (see Fig. \ref{fig2}). Instead, for lower
values of $n_e$, $m^*/m$ increases up to a maximum and eventually
decreases again as $n_e\rightarrow 0$. Higher values of $E_0$
emphasize the same trend, with higher and broader maxima of $m^*/m$
as $E_0$ increases.

The results plotted in Fig. \ref{fig3} clearly show how the
underlying diverging DOS, Eq.(\ref{DOS}), for $E_0\neq 0$ is
responsible for the enhancement of the effective mass. By reading
off from Fig. \ref{fig2} the values of $\mu$ corresponding to the
density values for which $m^*/m$ deviates from $1+\lambda$, it is
easy to realize that the enhancement of $m^*/m$ starts when $\mu$
becomes lower than $\sim E_0$, that is when the (bare) DOS diverges
as $\sqrt{E_0/E}$. In this situation, the coupling to the phonons is
no longer parametrized by $\lambda$ alone, but rather by an
effective coupling which takes into account the strongly varying DOS
at low energies.\cite{ciuchi1} As a matter of fact, for small
$\lambda$, by enhancing $E_0$ the system crosses over from a weak to
a strong coupling regime, where the mass enhancement can be
considerably larger than unity. It becomes therefore natural to
consider signatures of such SO induced strong el-ph coupling regime
also in the spectral properties of the electrons, which can provide
valuable information testable by tunneling and/or photoemission
experiments.\cite{gayone,hofmann,kroger}

\section{spectral properties}
\label{spectral}

\begin{figure}[t]
\protect
\includegraphics[scale=0.55,clip=true]{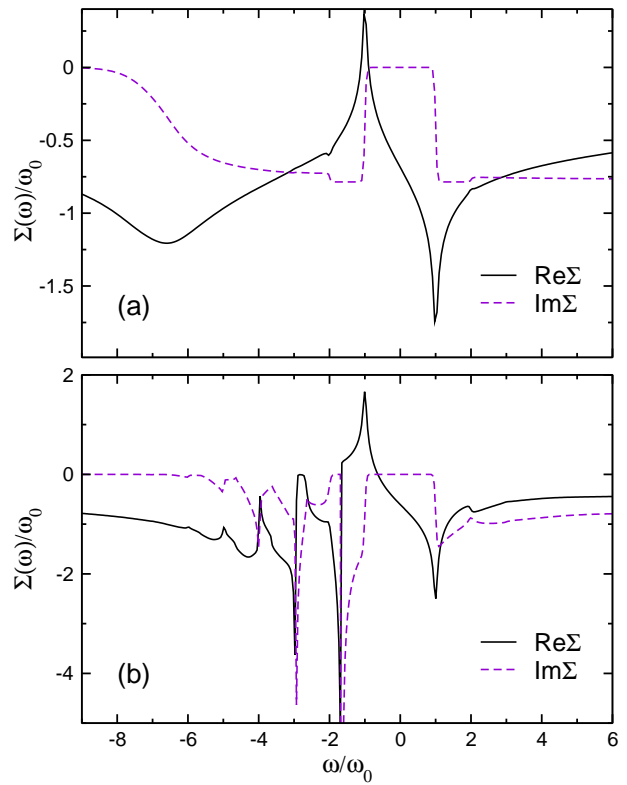}
\caption{(color online). Real and imaginary parts of the electron self-energy for $\lambda=0.5$
and electron density $n_e=0.1$. The SO energy is $E_0=0$ in (a)
and $E_0=5\omega_0$ in (b). } \label{fig4}
\end{figure}

The self-energy for real frequencies could be obtained directly from analytical
continuation on the real axis of Eqs. (\ref{self5}), (\ref{self6}) and (\ref{self7}).
However, since convergence is faster on the imaginary axis, in this paper
we opt for the more efficient method of analytical continuation formulated in
Ref.[\onlinecite{msc}]. Hence, once $\Sigma(i\omega_n)$ has been determined from
the imaginary axis equations (\ref{self5})-(\ref{self7}), the
retarded self-energy $\Sigma_R(\omega)=\Sigma(\omega+i\delta)$ follows from
\begin{eqnarray}
\label{selfreal}
\Sigma_R(\omega)&=&-T\lambda\sum_m D(\omega-i\omega_m)g(i\omega_m) \nonumber \\
&&+\lambda\frac{\omega_0}{2}[n(\omega_0)+f(\omega_0-\omega)]g_R(\omega-\omega_0) \nonumber \\
&&+\lambda\frac{\omega_0}{2}[n(\omega_0)+f(\omega_0+\omega)]g_R(\omega+\omega_0), \nonumber \\
\end{eqnarray}
where $g_R(\omega\pm\omega_0)=g(\omega\pm\omega_0 + i\delta)$ and $n(x)$ and $f(x)$ are
the distribution functions for bosons and fermions, respectively.
The real and imaginary parts of $\Sigma_R(\omega)$ are plotted in Fig. \ref{fig4}
for $\lambda=0.5$, $n_e=0.1$,  $E_0=0$ [Fig. \ref{fig4}(a)], and $E_0=5\omega_0$
[Fig. \ref{fig4}(b)]. The mass enhancement extracted from
$m^*/m=\lim_{\omega\rightarrow 0}[1-d{\rm Re}\Sigma_R(\omega)/d\omega]$ is
$1.45$ for $E_0=0$ and $1.72$ for $E_0=5\omega_0$, which agree with the $m^*/m$ values
plotted in Fig. \ref{fig3}. For $E_0=0$, the self-energy displays features typical
of the Holstein model for a 2D system in the non-crossing approximation.
Namely, ${\rm Im}\Sigma_R(\omega)=0$ for $\vert\omega\vert <\omega_0$,
while at larger frequencies ${\rm Im}\Sigma_R(\omega)\simeq -\pi\lambda\omega_0/2$.
The rapid decrease of $\vert{\rm Im}\Sigma_R(\omega)\vert$ at negative frequencies
stems from the bottom band edge. For $E_0=5\omega_0$ [Fig. \ref{fig4}(b)] the structure of
$\Sigma_R(\omega)$ is more intricate due to the strong energy dependence of the
underlying bare DOS. In fact, for $n_e=0.1$, the value of the (bare) chemical potential $\mu$
is well below $E_0=5\omega_0$ (see Fig. \ref{fig2}), and the $\omega$-dependence of
$\Sigma_R(\omega)$ becomes strongly influenced by the square-root divergence of the DOS.
This is particularly clear in Fig. \ref{fig4}(b), where ${\rm Im}\Sigma(\omega)$ reproduces
for $\omega<0$ the low-energy profile of the DOS shifted by multiples of $\omega_0$.
This feature is characteristic of a strongly-coupled el-ph system and
is fully consistent with the high value of the mass enhancement ($m^*/m\simeq 1.72$)
for this particular case.

\begin{figure*}
\protect
\includegraphics[scale=0.7,clip=true]{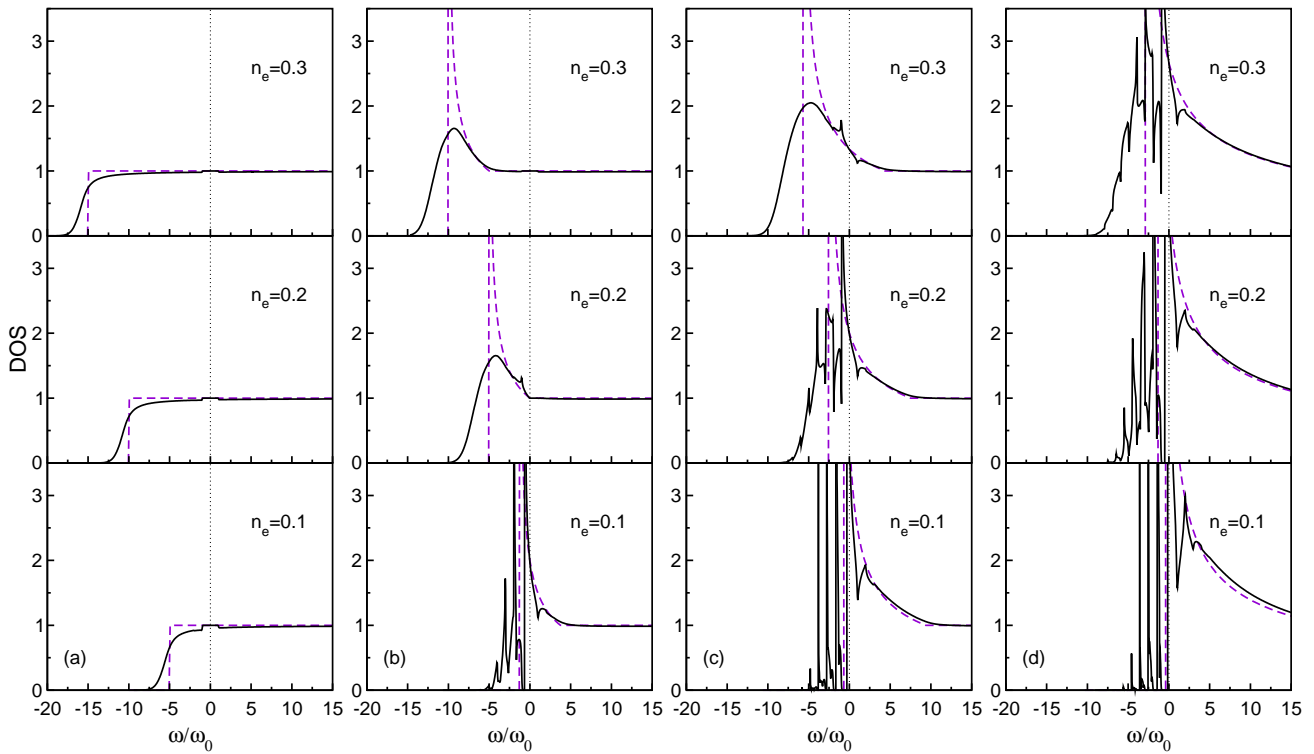}
\caption{(color online). Reduced DOS for $\lambda=0.5$  (solid lines)
and $\lambda=0$ (dashed lines). The vertical dotted line at $\omega=0$
indicates the Fermi level. The cutoff energy is $E_c=100\omega_0$ and the temperature
is $T=0.02\omega_0$ for the interacting cases ($T=0$ for $\lambda=0$).
(a): $E_0=0$, (b): $E_0=5\omega_0$, (c): $E_0=10\omega_0$, (d): $E_0=20\omega_0$.}
\label{fig5}
\end{figure*}

A global view of the behavior for several values of the electron number density and
of the SO energy is given in Fig. \ref{fig5} where the reduced DOS for the interacting
system
\begin{equation}
\label{DOSint}
\rho(\omega)=-\frac{1}{\pi}{\rm Im}g_R(\omega)
\end{equation}
is plotted for fixed $\lambda=0.5$. For comparison, we report also the
bare DOS $\rho_0(\omega)$, Eq.(\ref{DOS}), for the corresponding values of $n_e$ and $E_0$.
For $E_0=0$, Fig. \ref{fig5}(a), reducing the electron
density merely shifts the Fermi level for the interacting electron (vertical dotted line)
towards the bottom of the band. For $\vert\omega\vert <\omega_0$ $\rho(\omega)$ coincides
with the bare reduced DOS $\rho_0(\omega)=1$ because, as also shown in Fig. \ref{fig4},
the imaginary part of the self-energy is zero in that frequency range. Compared to the $\lambda=0$
case, whose DOS has a finite step at the bottom of the band, the profile of $\rho(\omega)$
is smeared by the el-ph interaction. A similar feature is obtained also for
$E_0=5\omega_0$ [Fig. \ref{fig5}(b)] and $n_e=0.3$ where, now, the square-root divergence
of $\rho_0(\omega)$ is rounded-off in $\rho(\omega)$ due to the finite lifetime
for $\lambda=0.5$. However, contrary to the $E_0=0$ case, reducing $n_e$ does not translate
to a (more or less) rigid shift of the Fermi level but, rather, creates new structures whose
intensity increases as the Fermi level moves deeper into the square-root singularity of $\rho_0(\omega)$.
This is even more pronounced for $E_0=10\omega_0$ and $E_0=20\omega_0$ plotted respectively in
Figs. \ref{fig5} (c) and (d). For the latter cases, the profile of $\rho(\omega)$ for $n_e=0.1$
is characterized by well defined peaks separated by multiples of $\omega_0$, and whose widths decrease
as $E_0/\omega_0$ is enhanced.

\begin{figure}[t]
\protect
\includegraphics[scale=0.65,clip=true]{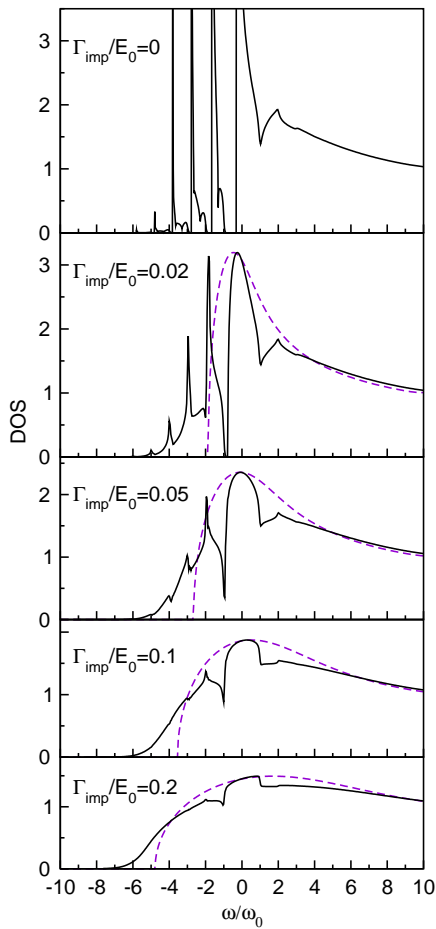}
\caption{(color online). Reduced DOS for $\lambda=0.5$ (solid lines) and $\lambda=0$
(dashed lines) for several values of the impurity scattering rate $\Gamma_{\rm imp}$.
The temperature is $T=0.02\omega_0$ and $E_0=10\omega_0$, $n_e=0.1$ for all cases.} \label{fig6}
\end{figure}

Such strong-coupling features are in principle directly observable by means of low-temperature
tunneling or photoemission measurements provided, however, that other interactions do
not alter significantly the profile of $\rho(\omega)$. This could be not the case for example
when disorder effects are taken into account, since these tend to smear all sharp features
of the DOS even at zero temperature. To investigate this point, we have considered a coupling
to a short-range impurity potential of the form
$V({\bf r})=V_{\rm imp}\sum_i\delta({\bf r}-{\bf R}_i)$, where ${\bf R}_i$ are the random positions of
impurity scatterers. Within the self-consistent Born approximation, the resulting self-energy is hence
given by Eq.(\ref{self6}) with the impurity term $\Gamma_{\rm imp}g(i\omega_n)/\pi$
added in the right-hand side, and Eq.(\ref{selfreal}) modified accordingly.
The parameter $\Gamma_{\rm imp}=1/2\tau_{\rm imp}=\pi n_iV_{\rm imp}^2 N_0$
is the usual scattering rate for zero SO coupling and for density $n_i$ of impurities.
In Fig. \ref{fig6} we report the calculated reduced DOS $\rho(\omega)$ for $\lambda=0.5$, $E_0=10\omega_0$,
$n_e=0.1$ and for several values of $\Gamma_{\rm imp}$. Also plotted by dashed lines
for $\Gamma_{\rm imp}\neq 0$
are the corresponding DOS curves in the absence of el-ph interaction ($\lambda=0$).
Compared to the clean limit $\Gamma_{\rm imp}=0$ (top panel), for rather weak disorder
($\Gamma_{\rm imp}=0.02 E_0$) the phonon peaks at $\omega<0$ are considerably less
sharp and slighly shifted at more negative frequencies, but nevertheless still clearly
discernible. Larger values of $\Gamma_{\rm imp}$ increasingly smear the phonon
structures, and gradually the peaks disappear. For $\Gamma_{\rm imp}=0.2 E_0$
(which corresponds to $\Gamma_{\rm imp}=2\omega_0$)
the resulting DOS is basically dominated by the impurity interaction, and does not deviate much
from the $\lambda=0$ case.

\section{discussion}
\label{disc} In this section we review the meaning of the
approximations used in the present work and discuss alternative
models for the study of the el-ph interaction in strong Rashba SO
systems. Let us start by considering the limitations of the
non-crossing approximation for the electron self-energy. For zero SO
interaction, or for Fermi energies sufficiently larger than $E_0$,
this is a rather good description of the el-ph problem provided
$\lambda$ is sufficiently small. However, as we have seen, for
non-zero SO interaction the electrons behave as effectively strongly
coupled to the phonons when the Fermi level is below $E_0$. This is
because of the square-root singularity in the DOS when $E_0\neq 0$.
In the $E_F< E_0$ regime, therefore, the non-crossing approximation,
although making evident the trend towards strong coupling, may not
be adequate for a more trustworthy description of the system. It is
instructive at this point to consider the limiting situation where
only one electron is present. By using second-order perturbation
theory (that is simply the non-crossing approximation with the bare
electron Green's function) it is easy to evaluate the mass
enhancement factor. At zero temperature, and setting for simplicity
$E_c/E_0=\infty$, this is given by
\begin{equation}
\label{2pt}
\frac{m^*}{m}=1+\frac{\lambda}{2}+
\frac{\lambda}{2}\sqrt{\frac{E_0}{\omega_0}}
\arctan\!\left(\sqrt{\frac{E_0}{\omega_0}}\right).
\end{equation}
It is clear from the above expression that for $E_0\gg\omega_0$ the mass
enhancement factor is governed by an effective coupling, say $\tilde{\lambda}$, proportional to
$\lambda\sqrt{E_0/\omega_0}$, amplified with respect to $\lambda$ by the
square-root divergent DOS. Equation (\ref{2pt}) clarifies also that the relevant
adiabatic parameter for $E_F\ll E_0$ is $\omega_0/E_0$, rather than $\omega_0/E_c$
(where $E_c$ plays the role of the bandwidth),
and that the effective coupling $\tilde{\lambda}$ increases as $\omega_0/E_0\rightarrow 0$.
Consequently, in the adiabatic limit $\omega_0/E_0= 0$ perturbation theory
breaks down for any finite $\lambda$ because $\tilde{\lambda}=\infty$.
This leads us to suspect that, in analogy
with the adiabatic limit of the one-dimensional lattice Holstein model,\cite{kabanov}
the ground state of a single electron for $\omega_0/E_0= 0$ is always a bound polaron.
However, for $\omega_0/E_0\neq 0$, the effective coupling
$\tilde{\lambda}\simeq \lambda\sqrt{E_0/\omega_0}$ is finite, which permits to estimate
a rough range of validity of Eq.(\ref{2pt}). Indeed, contributions of higher orders of
perturbation theory become negligible as long as $\tilde{\lambda}\ll 1$, corresponding to
$\lambda\ll \sqrt{\omega_0/E_0}$, consistent with the results on the one-dimensional
Holstein model of Refs.[\onlinecite{marsiglio,ccg}], which show better agreement
between perturbation theory and exact diagonalization results as $\omega_0$ increases.

Also for low but finite electron densities it is possible to
interpret the coupling to the phonons in terms of an
effective el-ph coupling $\tilde{\lambda}$ which grows as $E_0$ increases.
For example, in the range of
electron densities $n_e=0.1$-$0.2$, the mass enhancement factor
plotted in Fig. \ref{fig3} for $\lambda=0.5$ may be interpreted by
an effective Migdal-Eliashberg formula $m^*/m=1+\tilde{\lambda}$,
where $\tilde{\lambda}\approx 1$ for $E_0=10\omega_0$ and
$\tilde{\lambda}\approx 1.3$ for $E_0=20\omega_0$. However, contrary
to the one-electron case discussed above, now $\tilde{\lambda}$ depends
on the Fermi energy $E_F$. Indeed, provided that $\omega_0<E_F<E_0$, the effective
coupling turns out to be of order $\tilde{\lambda}\simeq\lambda\sqrt{E_0/E_F}$,
where the square-root term stems from the singularity of the DOS, Eq.(\ref{DOS}),
in analogy with the general definition of the effective el-ph interaction
in the presence of a van Hove singularity.\cite{cappelluti}
At this point it is possible to estimate the validity of the self-consistent
non-crossing approximation for the self-energy considered in the
previous sections. In fact, according
to Migdal's theorem generalized to systems with diverging DOS,\cite{cappelluti}
the el-ph vertex correction factors beyond the
non-crossing approximation are at least of order
$P=\tilde{\lambda}\omega_0/E_F$,
so that neglecting them would introduce an error of order $P$.
Estimates of $P$ for the different cases discussed in this paper can be
obtained by evaluating $\tilde{\lambda}\simeq 1-m^*/m$ from Fig.\ref{fig3}.
In this way the Fermi energy is roughly given by 
$E_F\simeq (\lambda/\tilde{\lambda})^2E_0$, which then can be inserted in 
the definition of $P$. For the low density value
$n_e=0.1$ we find $P\gtrsim 1$ for $E_0=20\omega_0$ and $P\simeq 0.5$
for $E_0=10\omega_0$, showing that the non-crossing approximation is
quantitatively inaccurate in this case.
However, already for $n_e=0.2$, for which effectively
strong-coupling features are apparent from Fig.\ref{fig3} and Fig.\ref{fig4},
the contributions of the vertex corrections drop to $P\simeq 0.4$ and $P\simeq 0.2$
for $E_0=20\omega_0$ and $E_0=10\omega_0$, respectively. In this situation,
the non-crossing approximation is fairly reliable and its accuracy improves
as $n_e$ is further enhanced and/or $E_0/\omega_0$ is reduced.

Let us turn now to discuss the general form of the self-energy
for the case in which the el-ph matrix element is momentum dependent.
Here we consider the situation in which the momentum dependence is
only through the modulus $q$ of the momentum transfer ${\bf q}$,
as is the case, for example, with the 2D Fr\"ohlich model, for
which the coupling goes like $1/\sqrt{q}$. As shown in the Appendix,
a fully general expression of the self-energy valid also beyond
the non-crossing approximation is:
\begin{equation}
\label{selfq}
\mathbf{\Sigma}({\bf k},i\omega_n)=\Sigma_1(k,i\omega_n)\mathbf{1}+\Sigma_2(k,i\omega_n)
\hat{\mathbf{\Omega}}_{\bf k}\cdot\mbox{\boldmath$\sigma$},
\end{equation}
where $\Sigma_1$ and $\Sigma_2$ are scalars. Compared to Eq.(\ref{self6}), the
above expression has an additional term which is off-diagonal in the spin subspace,
renormalizing therefore the SO coupling. This term disappears ($\Sigma_2=0$)
only when the el-ph matrix element is momentum independent, like in the Holstein model,
and at the same time the self-energy is evaluated in the non-crossing approximation.
In all other cases, like, {\it e.g.}, the Fr\"ohlich model in the non-crossing
approximation, $\Sigma_2$ is nonzero.
For sufficiently large values of $E_F,$ such that the weak SO limit $E_0/E_F\ll 1$ holds true
so that the Fermi level lies far above the 1D singularity of the DOS,
$\Sigma_2$ turns out to be of order
$\lambda\omega\Delta_{\rm so}/E_F\propto\lambda\omega_0\sqrt{E_0/E_F}$, and can be disregarded
in comparison with $\Sigma_1\approx\lambda\omega_0$. On the contrary, when
$E_F/E_0\lesssim 1$, $\Sigma_1$ and $\Sigma_2$ have comparable magnitude, and
the full momentum and frequency dependent of both terms must be considered for a consistent
evaluation of the el-ph effects.

\section{conclusions}
\label{concl} In this paper we have addressed the role of the Rashba
SO interaction in the properties of a coupled el-ph gas in two
dimensions. By using a self-consistent non-crossing approximation
for the electron self-energy, we have studied the mass enhancement
factor and the spectral properties. We have shown that, for
sufficiently strong SO interaction, the electron becomes strongly
coupled to the phonons even if el-ph coupling $\lambda$ can be
classified as weak. We identify this behavior as being due to a
topological change of the Fermi surface for strong SO interaction,
which gives rise to a square-root singularity in the DOS at low
energies. Signatures of such effectively strong el-ph coupling are
found in the mass enhancement factor, which becomes as large as
$m^*/m\approx 2$ for el-ph coupling of only $\lambda=0.5$, and in
the energy dependence of the interacting DOS, displaying low energy
peaks separated by multiples of the phonon energy $\omega_0$.
This latter feature could be tested experimentally by tunneling or
photoemission experiments in systems where the Fermi level can
be tuned to approach the square-root singularity of the DOS. We
have then discussed limitations of the non-crossing approximation
approach and possible generalizations of the theory for
momentum-dependent el-ph matrix elements. Since the problem of el-ph
coupling in the presence of SO interaction is relevant for several
systems such as metal and semimetal surface states, surface
superconductors, or low-dimensional heterostructures, and given the
current interest in spintronic physics, we hope that our work will
stimulate further investigations.

\appendix*
\section{}
\label{appe}
In this appendix we evaluate the form of the electron self-energy
when the el-ph interaction is momentum dependent. In particular,
we consider the hamiltonian $H=H_0+H_{ph}$, where $H_0$ is the
Rashba spin-orbit hamiltonian of Eq.(\ref{ham}) and
\begin{equation}
\label{app1}
H_{ph}= \sum_{\bf q}\omega_0a^\dagger_{\bf q}a_{\bf q}+
\sum_{\mathbf{k}\mathbf{k}'\alpha}g_{{\bf k}-{\bf k}'}
c^\dagger_{\mathbf{k}\alpha}c_{\mathbf{k}'\alpha}
(a^\dagger_{\mathbf{k}-\mathbf{k}'}+a_{\mathbf{k}'-\mathbf{k}}),
\end{equation}
where $g_{\bf q}$ is the el-ph matrix element which we assume
depends only on the modulus of momentum transfer ${\bf q}$.
It is convenient to rewrite $H$ in terms
of the eigenvectors of $H_0$, whose annihilation operators $d_{{\bf k}s}$
($s=\pm$) are related to $c_{{\bf k}\alpha}$ through
\begin{equation}
\label{app2}\left(
\begin{array}{c}
d_{{\bf k}+} \\
d_{{\bf k}-}
\end{array}\right)=\mathbf{T}_{\bf k}
\left(
\begin{array}{c}
c_{{\bf k}\uparrow} \\
c_{{\bf k}\downarrow}
\end{array}\right)=\frac{1}{\sqrt{2}}\left(
\begin{array}{cc}
1 & -ie^{-i\varphi} \\
1 & ie^{-i\varphi}
\end{array}\right)
\left(
\begin{array}{c}
c_{{\bf k}\uparrow} \\
c_{{\bf k}\downarrow}
\end{array}\right),
\end{equation}
where $\varphi$ is the azimuthal angle of ${\bf k}$. In this basis, $H_0$
is diagonal with dispersion relation given by Eq.(\ref{energy2}) while
$H_{ph}$ becomes
\begin{equation}
\label{app3}
H_{ph}= \sum_{\bf q}\omega_0a^\dagger_{\bf q}a_{\bf q}+
\sum_{\mathbf{k}\mathbf{k}'ss'}M^{s,s'}_{{\bf k}-{\bf k}'}
d^\dagger_{\mathbf{k}s}d_{\mathbf{k}'s'}
(a^\dagger_{\mathbf{k}-\mathbf{k}'}+a_{\mathbf{k}'-\mathbf{k}}),
\end{equation}
with
\begin{equation}
\label{app4}
M^{s,s'}_{{\bf k}-{\bf k}'}=g_{{\bf k}-{\bf k}'}
\frac{1+ss'e^{i(\varphi-\varphi')}}{2}.
\end{equation}
By applying Wick's theorem, it turns out that,
to all orders of the el-ph interaction, the Green's function
in the chiral basis has zero off-diagonal components so that,
if $G_\pm({\bf k},\tau)=-\langle T_\tau d_{{\bf k}\pm}(\tau)
d^\dagger_{{\bf k}\pm}(0)\rangle$,
the matrix Green's function in the original spin sub-space becomes
\begin{align}
\label{app5}
\mathbf{G}({\bf k},i\omega_n)&=\mathbf{T}^\dagger_{\bf k}
\left(
\begin{array}{cc}
G_+({\bf k},i\omega_n) & 0 \\
0 & G_-({\bf k},i\omega_n)
\end{array}
\right)\mathbf{T}_{\bf k} \nonumber \\
&=\frac{1}{2}\sum_{s=\pm}(1+s\hat{\mathbf{\Omega}}_{\bf
k}\cdot\mbox{\boldmath$\sigma$})G_s({\bf k},i\omega_n).
\end{align}
Consequently, by using Dyson's equation (\ref{self1}), the matrix self-energy
in the spin sub-space is
\begin{equation}
\label{app6}
\mathbf{\Sigma}({\bf k},i\omega_n)=\Sigma_1({\bf k},i\omega_n)\mathbf{1}
+\Sigma_2({\bf k},i\omega_n)
\hat{\mathbf{\Omega}}_{\bf k}\cdot\mbox{\boldmath$\sigma$},
\end{equation}
where
\begin{equation}
\label{app7}
\Sigma_{1(2)}({\bf k},i\omega_n)=\frac{\Sigma_+({\bf k},i\omega_n)+(-)
\Sigma_-({\bf k},i\omega_n)}{2}.
\end{equation}
In order to obtain Eq.(\ref{selfq}), it suffices to demonstrate
that the momentum dependence of self-energy in the chiral basis,
$\Sigma_\pm({\bf k},i\omega_n)$, is only via $k=\vert{\bf k}\vert$.
This is accomplished by noticing that the el-ph matrix element in the
chiral basis, Eq.(\ref{app4}), depends on the direction of the momentum
transfer ${\bf k}-{\bf k}'$ solely through $\varphi-\varphi'$. Hence,
if the electronic dispersion depends only on the modulus of the momentum,
as is the case with Eq.(\ref{energy2}), a general self-energy diagram
in the chiral basis will be independent of the direction of ${\bf k}$
which, by using (\ref{app7}), is consistent with Eq.(\ref{selfq}).

\end{document}